\documentstyle[12pt,aasms4]{article}
\slugcomment{Submitted to The Astrophysical Journal}

\def\lsim{{{}_{{}_<}^{~}\atop {}^{{}^\sim}_{~}}}
\newcommand{\be}{\begin{equation}}
\newcommand{\ee}{\end{equation}}
\newcommand{\lab}[1]{\label{#1}}
\newcommand{\r}[1]{~(\ref{#1})}

\begin{document}

\title{The Zero Point of Extinction Toward Baade's Window From RR Lyrae Stars}

\author {C.  Alcock$^{1,2}$, R.A.  Allsman$^{1,4}$, D. R. Alves$^{1,10}$,
T.S. Axelrod$^{1,4}$, A.C. Becker$^{2,3}$, 
D.P. Bennett$^{1,2}$, K.H. Cook$^{1,2}$, 
K.C. Freeman$^4$, K. Griest$^{2,6,11}$, A. Gould$^{5,11}$, 
J.A. Guern$^6$, M.J. Lehner$^6$, 
S.L. Marshall$^{1}$, D. Minniti$^{1}$,
B.A. Peterson$^4$, P. Popowski$^{5,12}$
M.R. Pratt$^{2,3}$, P.J. Quinn$^7$, A.W. Rodgers$^4$, C.W.
Stubbs$^{2,3,7,11}$, W. Sutherland$^8$, T. Vandehei$^{6}$, 
and D.L. Welch$^{9}$}
\affil { (The MACHO Collaboration) }

\altaffiltext{1}{Lawrence Livermore National Laboratory, Livermore, CA 94550\\
E-mail:  alcock, robynallsman, tsa, bennett, kcook, dminniti, 
stuart@llnl.gov}

\altaffiltext{2}{Center for Particle Astrophysics, University of California,
Berkeley, CA 94720}

\altaffiltext{3}{Department of Astronomy, University of Washington,
Seattle, WA 98195 \\
E-mail: stubbs, becker@astro.washington.edu}

\altaffiltext{4}{{Mt~Stromlo}~\&~Siding~Spring~Observatories, 
Australian~National~University,\linebreak ~Weston~ACT~2611,~Australia \\ E-mail: kcf, peterson, 
pjq, alex@merlin.anu.edu.au}

\altaffiltext{5}{Department of Astronomy, Ohio State University,
Columbus, OH 43210 \\
E-mail: gould, popowski@astronomy.ohio-state.edu}

\altaffiltext{6}{Department of Physics, University of California,
San Diego, CA 92093 \\
E-mail: griest, jguern, matt, vandehei@astrophys.ucsd.edu}

\altaffiltext{7}{European Southern Observatory, D-85748 Garching bei 
M\"unchen, Garmany\\
E-mail: pjq@eso.org}

\altaffiltext{8}{Department of Physics, University of Oxford,
Oxford OX1 3RH, U.K.\\
E-mail: wjs@oxds02.astro.ox.ac.uk}

\altaffiltext{9}{Department of Physics and Astronomy, McMaster University,
Hamilton, ON L8S 4M1, Canada\\
E-mail: welch@physics.mcmaster.ca}

\altaffiltext{10}{Department of Physics, University of California, Davis, CA 
95616}

\altaffiltext{11}{Alfred P. Sloan Foundation Fellow}

\altaffiltext{12}{Work toward Ph.D thesis}

\begin{abstract}
	We measure the zero point of the Stanek (1996) extinction map
by comparing the observed $(V-K)$ colors of 20 RR Lyrae stars (type ab) found
in the MACHO survey with their intrinsic $(V-K)_0$ colors as a function of
period as determined from nearby RR Lyrae stars.  We find
that the zero point of the Stanek map should be changed by 
$\Delta A_V = -0.11\pm 0.05$ mag, in excellent agreement with the recent
measurement of Gould, Popowski \& Terndrup (1997) using K giants.

Subject Headings:  dust, extinction -- Galaxy: general -- stars: variables:
other
\end{abstract}

\section{Introduction}

	Baade's Window, $(\ell,b)\sim (1^\circ,-4^\circ)$, has been 
an important laboratory for the study of bulge populations because of its
relatively low extinction $(A_V\sim 1.5)$ and the presence of the globular
cluster NGC 6522 which 
provided an early opportunity to measure that extinction.  Baade's Window has
been a major focus of microlensing surveys by OGLE (Udalski et al.\ 1993),
MACHO (Alcock et al.\ 1995), and EROS (Ansari et al.\ 1996).
Among the many non-microlensing applications of these surveys is Stanek's
(1996) construction of a detailed extinction map with $30''$ resolution
over the $40'$ square OGLE field.  Stanek (1996) applied the method of 
Wo\'zniak \& Stanek (1996), in essence measuring the mean apparent  
magnitude, $\langle V\rangle$, and 
color, $\langle V-I\rangle$, of clump giants as a function of position.  He
then inferred from these the {\it differential} total and selective 
extinctions, $A_V$ and $E(V-I)$.  The estimated relative error in $A_V$ of
this map is $0.1\,$mag for individual resolution elements, but there is an 
overall uncertainty of $0.2\,$mag in the zero point.

	For many applications, such as the interpretation
of color-magnitude diagrams of bulge field stars and of the cluster NGC 6522
or the measurement of distances using RR Lyraes or other tracers, the 
determination of the zero point is crucial.  

\section{Previous Work}

\subsection{$E(B-V)$}	

Historically, the extinction toward Baade's Window has been found by 
measuring the selective extinction $E(B-V)$, and then multiplying by
an assumed ratio of total to selective extinction $R_V=A_V/E(B-V)$
(Arp 1965; van den Bergh 1971; Walker \& Mack 1986; Terndrup \& Walker 1994).
There are several major disadvantages to this approach.  First, $R_V\sim 3$
is rather large, and the statistical error in $E(B-V)$ (usually estimated
to be $\geq 0.03$) is multiplied by this factor when estimating the error
in $A_V$.  Second, $R_V$ varies along different lines of sight, so for any
particular line of sight for which it is not actually measured (e.g., Baade's
Window) the precision of the estimate is no better than 7\%.  Hence, the
statistical error alone for $A_V$ is more than 0.12 mag.  Finally, there
are systematic errors arising from uncertainties in the intrinsic $(B - V)$
colors of stars used to estimate $E(B-V)$.  While the intrinsic color of
extremely hot stars (in the Raleigh-Jeans limit) is known from fundamental 
physics, there are no known stars at such extreme temperatures
lying beyond the dust column in this direction.  Hence, one must use
cooler stars whose $(B - V)$ colors are sensitive functions of temperature,
metallicity, and
perhaps other factors.  The standard approach is to find local analogs of
the program stars and directly measure their colors, but systematic errors
may arise from any unrecognized differences between these two groups of stars.
As always, it is difficult to determine the size of the systematic errors, but
one can gain a sense of their
magnitude by comparing the $E(B-V)_0=0.45\pm 0.04$ derived
by van den Bergh (1971) from three different methods based on cool stars
(K and M giants) with the $E(B-V)_0=0.60\pm 0.03$ derived by Walker \& Mack
(1986) using relatively hot stars (RRab Lyraes).  Here the subscript ``0'' 
means  ``reduced to zero color'' using the prescription of Dean, Warren, \&
Cousins (1978). 

\subsection{$E(V-K)$}	

	Terndrup, Sadler, \& Rich (1995) pioneered a radically different 
approach.  They measured $E{(V-K)}=1.23\pm 0.08$
for a sub-region of Baade's Window (``Blanco region A'')
with relatively uniform extinction by comparing the 
$H\beta \, \lambda 4861$ (Faber et al. 1985) index as a function of 
$(V-K)$ color
to that observed for bright K giants in the solar neighborhood.
They then inferred (but did not explicitly write down),
$A_{V} = {E(V-K)/(1-\alpha)} = 1.38\pm 0.09$, where
\be 
\alpha \equiv {A_K\over A_V} = 0.112\pm 0.002, \lab{rkdef}
\ee
is the ratio of $K$ to $V$ extinction (Rieke \& Lebofsky 1985).
While this approach is formally identical to the previous one (measurement of
a selective extinction and conversion to a total extinction), it is 
potentially more accurate than using $E(B-V)$ because the extrapolation to 
total visual extinction is small (a factor 1.12 vs.\ 3),
and therefore the error in $A_V$
is only slightly bigger than the error in $E(V-K)$.

	Stanek (1996) set the zero point of his extinction map by forcing it
to reproduce the results of Terndrup et al.\ (1995) over their sub-region.
However, as discussed in some detail by Gould, Popowski, \& Terndrup (1997),
Stanek (1996) incorporated inconsistent assumptions in setting the zero point.
Gould et al.\ (1997) therefore estimated a ``naive'' correction to the
Stanek (1997) map of $\Delta A_V=-0.09\pm 0.09$.

	However, Gould et al.\ (1997) also recognized that the very existence
of the Stanek (1996) {\it differential} map makes possible a much more 
accurate determination of the {\it zero point}.  One can now use the
{\it individual} differential extinctions $A_{V,i}^{\rm Stanek}$ from the
Stanek (1996) map for an ensemble of stars $i=1...n$ to make individual
estimates for the correction to the Stanek (1996) map,
\be
\Delta A_{V,i} = {(V-K)_{i} - (V-K)_{0,i}\over 1-\alpha} - 
A_{V,i}^{\rm Stanek}. \lab{deltavmk}
\ee
Here $(V-K)_{i}$ is the observed color of the star and $(V-K)_{0,i}$
is its predicted dereddened color.  Gould et al.\ (1997) applied this
method to a sample of $n=206$ K giants from Terndrup et al.\ (1995) and
derived
\be
\Delta A_V = {\langle \Delta A_{V,i}\rangle}
=-0.10\pm 0.06\qquad {(\rm K}\ {\rm Giants}). \lab{deltaavgpt}
\ee
Here we apply equation\r{deltavmk} to a sample of $n=20$ RR Lyrae stars 
(type ab) found by the MACHO collaboration.  
RR Lyrae stars are substantially hotter than
K giants and the method by which we infer their dereddened $(V-K)_0$
colors is substantially different than that used by Gould et al.\ (1997).
Therefore, this new determination yields an important check on any
unrecognized systematic effects which may have affected previous results.

\section{Calibration}

	Figure 1 shows the dereddened $(V-K)_0$ colors for 20 RR Lyraes
plotted against the fundamentalized period, $P_0$, taken from Table 9 of
Jones et al.\ (1992).  For the 17 RRab's
({\it circles}), this quantity is the same as the observed period, while for 
the 3 RRc's ({\it crosses}) it is inferred from the observed first-overtone 
period $P_1$.  The solid line is the best fit to the RRab's and is given by
\be
(V-K)_0 = 1.046\pm 0.013 + (1.245\pm 0.144)(\log P_0 + 0.29),
\lab{vmknought}
\ee
where the expression gives the fit in a form with uncorrelated
errors.  In more traditional format, this becomes
$(V-K)_0 = 1.407 + 1.245\log P_0$.
It is also
possible to include [Fe/H] as an independent variable, in which case the
best fit is $(V-K)_0 = 1.650 + 1.797\log P_0 + 0.084\rm [Fe/H]$.  However,
as we show below, there is little advantage to this.

	It is clear from Figure 1 that the three RRc's do not follow the
same relation as the RRab's.  Moreover, there are not enough RRc's
to determine from the data the relation that is appropriate for them.  We
therefore exclude RRc's from further consideration.

%\placefigure{fig1}

\section{Data}

	The data are drawn from four sources and are presented in Table 1.
Columns 1 and 2 show the positions of each star in 2000 coordinates. Column 3
gives the periods and column 4 
gives the apparent $V$ magnitudes, both from MACHO data (see \S 4.1).  
Column 5 gives the apparent $K$ magnitudes from Table 1 of Carney et al.\ 
(1995) (see \S 4.2).  Column 6 gives the metallicity from Walker \& Terndrup
(1991).  Column 7 gives the visual extinction from the Stanek (1996) map.
Columns 8 and 9 gives the actual $(V-K)$ (from columns 4,5 and 7) and the 
$(V-K)$ predicted on the basis of equation\r{vmknought} using columns 3 and 7.

\subsection{$V$ Band Photometry}

	The apparent $V$ magnitudes listed in Table 1 are each based upon of 
order 100 observations taken by MACHO in each of two passbands $B_M$ and 
$R_M$, during the 1993 bulge season.  For each star, we find the magnitudes at 
mean flux, $\langle B_M\rangle$ and $\langle R_M\rangle$,
by taking the average flux (after excluding bad data points)
for all observations, and then converting to a magnitude.  Since the dispersion
of a typical RRab light curve is $\sim 0.3$ mag, this procedure produces a 
random  error of $\sim 0.03$ mag (relative to perfect coverage of the light
curve) which is small by comparison with other errors.
We convert these instrumental magnitudes to standard Johnson $V$ band using
$V = 23.67 + 1.0026 B_M  - 0.156(B_M-R_M)$ (Alves 1997). 
We apply this conversion to
the magnitudes at mean flux,  $\langle B_M\rangle$ and $\langle R_M\rangle$.

%\placefigure{fig2}

	As a check on the zero point of this conversion, we compare in
Figure 2 the $V$ magnitudes at mean flux derived from MACHO data with
those derived by Olech (1997) from 
OGLE data for 50 RRab's in common
between the two data sets.  Since Olech (1997) reports mean magnitudes, not
magnitudes at mean flux, we first correct the OGLE values by the difference
between the two quantities as determined from the MACHO data.  The offset of
OGLE relative to MACHO is $0.006\pm 0.019$ mag, with a scatter of 0.13 mag
in the difference.  That is, the zero points are in excellent agreement.
We note that Alcock et al.\ (1997) found a scatter of 0.12 mag in the mean
mags determined from MACHO photometry of individual bulge RR Lyrae stars by 
comparing 44 stars detected in overlapping fields.  Hence, the scatter in
the MACHO/OGLE difference can largely be accounted for by the MACHO scatter.

\subsection{$K$ Band Photometry}

	The apparent $K$ magnitudes listed in Table 1 are taken from Carney
et al.\ (1995).  Generally, these are based on observations at a single 
epoch.  Nevertheless, because the ephemerides were well known and because, in
any event, the amplitudes of RRab's in $K$ band are only $\lsim 0.3$ mag,
Carney et al.\ (1995) believe that the individual photometry errors are only
$\sim 0.03$ mag.

\section{Determination of the Zero Point}

	Figure 3 shows the individual estimates for $\Delta A_{V,i}$ for
the 20 stars for which data are available.  Note that $\Delta A_{V,i}$ 
is $(1-\alpha)^{-1}$ times the difference between
the last two columns in Table 1 (see eq.\r{deltavmk}).  The mean value is
$\langle{\Delta A_{V}}\rangle = -0.11$ with a scatter of 0.22 mag.  The best
estimate of the offset to the Stanek (1996) map is therefore,
\be
\Delta A_V = -0.11\pm 0.05\qquad {(\rm RR}\ {\rm Lyraes}).\lab{deltaavrr}
\ee

	The error in this estimate is determined from the scatter.  
In principle, there are also uncertainties in the zero points of MACHO 
photometry and of equation\r{vmknought}.  However, from 
the discussions in \S 3 and \S 4.1, we conclude that these zero-point
errors are negligible compared to equation\r{deltaavrr}.

%\placefigure{fig3}

	As discussed in \S 3, it is also possible to incorporate metallicity
information when predicting $(V-K)$ colors.  If we do so, however, we find 
that we recover equation\r{deltaavrr} exactly.  That is,
metallicity contains
no additional information over and above that
 already contained in the period.  This may
well be because the errors in the Walker \& Terndrup (1991) metallicity 
measurements are larger than the scatter in the period-color relation shown
in Figure 1.  In any event, since metallicity information does not 
improve (or change) the determination, we ignore it.

	The scatter in Figure 3 can be roughly accounted for as 
follows.  There is a scatter of 0.06 mag in $(V-K)$ from the calibrating 
relation\r{vmknought}.  
The Stanek (1996) map has an error of $0.10(1-\alpha)=0.09$ in 
$E(V-K)$.  The error in $K$ and $V$ photometry are 0.03 mag and 0.12 mag,
respectively (see \S 4).  Adding
these values in quadrature and multiplying by $(1-\alpha)^{-1}$ yields a 
predicted scatter of 0.18 mag in $\Delta A_{V,i}$ compared to 
$0.22\pm 0.04$ mag actually observed.

\section{Discussion}

	Equation\r{deltaavrr} based on RR Lyrae stars (type ab) agrees very 
well with equation\r{deltaavgpt} based K giants.  This is heartening because
the largest discrepancy among all methods of determining $E(B-V)_0$ was
between RR Lyraes, $E(B-V)_0=0.60\pm 0.03$ derived by Walker \& Mack (1986),
and cool giants, $E(B-V)_0=0.45\pm 0.04$ derived
by van den Bergh (1971).  The same discrepancy is not present when similar
classes of stars are used to determine the extinction via measurement of
$E(V-K)$.  We conclude that the $E(V-K)$ method has passed an important
test for systematic errors.  We therefore combine the two determinations to
obtain,
\be
\Delta A_V = -0.11\pm 0.04\qquad {(\rm combined}),\lab{deltaavall}
\ee
as our best overall estimate.  This corresponds to $A_V=1.36$ for the
``Blanco A region'' originally used by Stanek (1996) to calibrate his map.

	Our approach does not permit a direct measurement of $E(B-V)_0$ for
this region.  However, if we follow Terndrup et al.\ (1995) and adopt
$A_V/E(B-V)_0=2.85$, we derive $E(B-V)_0\sim 0.48$ which is closer to the
value derived by van den Bergh (1971) from cool giants than it is to the value
derived by Walker \& Mack (1986) from RR Lyrae stars.

\acknowledgements  
We are very grateful for the skilled support by the technical staff at Mount
Stromlo Observatory.
Work at LLNL is supported by DOE contract W7405-ENG-48.
Work at the CfPA is supported NSF AST-8809616 and AST-9120005.
Work at MSSSO is supported by the Australian Department of Industry,
Technology and Regional Development.
Work at OSU is supported in part by grant AST 94-20746 from the NSF.
WJS is supported by a PPARC Advanced Fellowship.
KG thanks support from DOE OJI, Sloan, and Cottrell awards.
CWS thanks support from the Sloan, Packard and Seaver Foundations.

\clearpage

\begin{table*}
{\scriptsize
\begin{center}
\begin{tabular}{rrrcrrrcccccccc}
\tableline\tableline
\multicolumn{3}{c}{$\alpha$} && \multicolumn{3}{c}{$\delta$} && $P_0$ & $V$ & $K$ & [Fe/H] & $A_V$ & $(V-K)_{obs}$ & $(V-K)_{pred}$ \\
\tableline
18 &3 &52.8036 && -29 &51 &46.326 && 0.4632 & 16.82 & 14.26 & -0.75 & 1.67 & 1.07 & 0.99 \\
18 &4  &4.2705 && -29 &56 &44.909 && 0.5158 & 16.74 & 14.29 & -0.91 & 1.56 & 1.06 & 1.05 \\
18 &2 &44.9499 && -29 &53  &8.325 && 0.6002 & 16.93 & 14.48 & -0.64 & 1.64 & 0.99 & 1.13 \\
18 &2 &38.9884 && -29 &51 &56.299 && 0.5296 & 16.88 & 14.52 & -0.95 & 1.63 & 0.91 & 1.06 \\
18 &3  &0.5605 && -29 &53 &51.781 && 0.5545 & 16.34 & 14.07 & -1.43 & 1.53 & 0.91 & 1.09 \\
18 &2 &57.6507 && -29 &48 &39.167 && 0.4241 & 16.70 & 14.50 & -0.94 & 1.43 & 0.93 & 0.94 \\
18 &3  &7.7948 && -29 &50  &5.678 && 0.5772 & 16.54 & 14.23 & -1.19 & 1.39 & 1.07 & 1.11 \\
18 &2 &39.1933 && -30  &7  &7.298 && 0.5945 & 17.22 & 14.25 & -0.56 & 1.84 & 1.33 & 1.13 \\
18 &2 &32.4351 && -29 &58 &56.447 && 0.4950 & 16.82 & 14.40 & -0.89 & 1.50 & 1.08 & 1.03 \\
18 &2 &54.9181 && -29 &59 &57.477 && 0.4786 & 16.95 & 14.49 & -0.42 & 1.50 & 1.12 & 1.01 \\
18 &3 &55.9506 && -30 &12 &45.772 && 0.4574 & 16.84 & 14.40 & -1.65 & 1.64 & 0.98 & 0.98 \\
18 &3 &39.8186 && -30  &9  &0.611 && 0.4897 & 16.93 & 14.67 & -0.84 & 1.70 & 0.75 & 1.02 \\
18 &4 &19.5733 && -29 &58 &26.145 && 0.5071 & 16.41 & 14.32 & -1.16 & 1.65 & 0.62 & 1.04 \\
18 &3 &18.6581 && -30  &1  &9.076 && 0.4543 & 16.85 & 14.76 & -1.35 & 1.45 & 0.80 & 0.98 \\
18 &3 &23.2661 && -30  &2 &47.075 && 0.4402 & 16.65 & 14.84 & -1.36 & 1.47 & 0.50 & 0.96 \\
18 &3 &18.6445 && -30  &5 &51.031 && 0.5715 & 16.86 & 14.18 & -1.41 & 1.69 & 1.18 & 1.10 \\
18 &1 &56.0269 && -29 &59 &55.655 && 0.4600 & 16.30 & 13.39 & -1.05 & 1.99 & 1.14 & 0.99 \\
18 &4 &29.4679 && -30  &1  &7.034 && 0.4970 & 16.71 & 14.53 & -1.03 & 1.52 & 0.83 & 1.03 \\
18 &3  &9.5146 && -30 &11 &56.964 && 0.6516 & 16.94 & 14.52 & -0.32 & 1.69 & 0.92 & 1.18 \\
18 &3 &57.2100 && -30  &6  &5.300 && 0.7705 & 16.64 & 14.30 & -1.30 & 1.62 & 0.90 & 1.27 \\
\tableline
\end{tabular}
\end{center}
}
\tablenum{1}
\caption{Columns 1 and 2 show the positions of each star in 2000 coordinates. Column 3
gives the period and column 4 
gives the apparent $V$ magnitudes, both from MACHO data.  
Column 5 gives the apparent $K$ magnitudes from Carney et al.\ 
(1995).  Column 6 gives the metalicity from Walker \& Terndrup
(1991).  Column 7 gives the visual extinction from the Stanek (1996) map.
Columns 8 and 9 gives the actual $(V-K)$ (from columns 4,5 and 7) and the 
$(V-K)$ predicted.}
\end{table*}

\clearpage

\figcaption[fig1.ps]{Period-color relation for 20 nearby RR Lyrae stars including 17 RRab's 
({\it circles}) and 3 RRc's ({\it crosses}).  Data are taken from Jones et al.\
(1992).  Straight line is the best fit to the RRab's:
$(V-K)_0 = 1.407 + 1.245\log P_0$.  Note that RRc's do not follow this 
relation.  They are therefore excluded from further consideration. \label{fig1}}

\figcaption[fig.ps]{Difference between MACHO and OGLE photometry 
$(V_{MACHO} - V_{OGLE})$ for 50 RRab's as a function of MACHO magnitude, 
$V_{MACHO}$.  The OGLE data are from Olech (1997) and have been corrected 
from mean magnitude to magnitude at mean flux.  The mean offset is 
$\langle V_{MACHO} - V_{OGLE}\rangle = 0.006 \pm 0.019$. \label{fig2}}

\figcaption[fig3.ps]{Individual estimates of the zero-point offset to the Stanek (1996) map,
$\Delta A_{V,i}$ for 20 RRab's.  See Table 1 and equation\r{deltavmk}.  The
mean offset is $\Delta A_V = -0.11\pm 0.05$.
\label{fig3}}

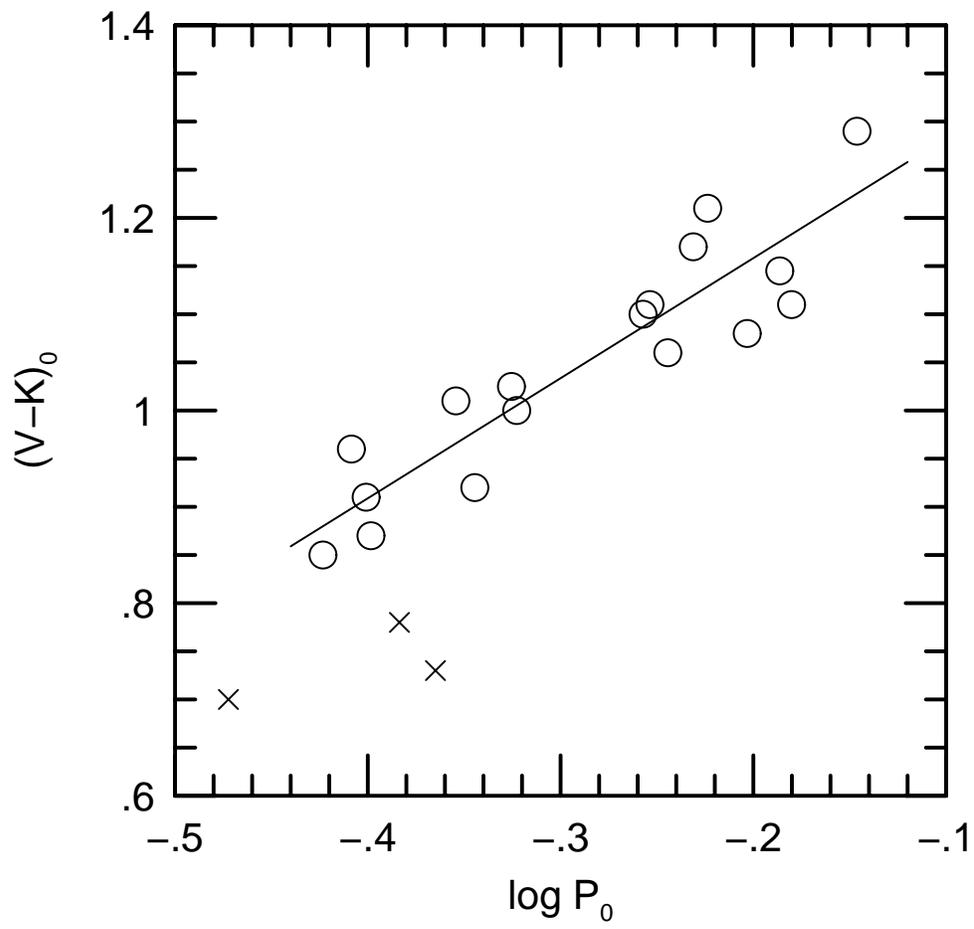
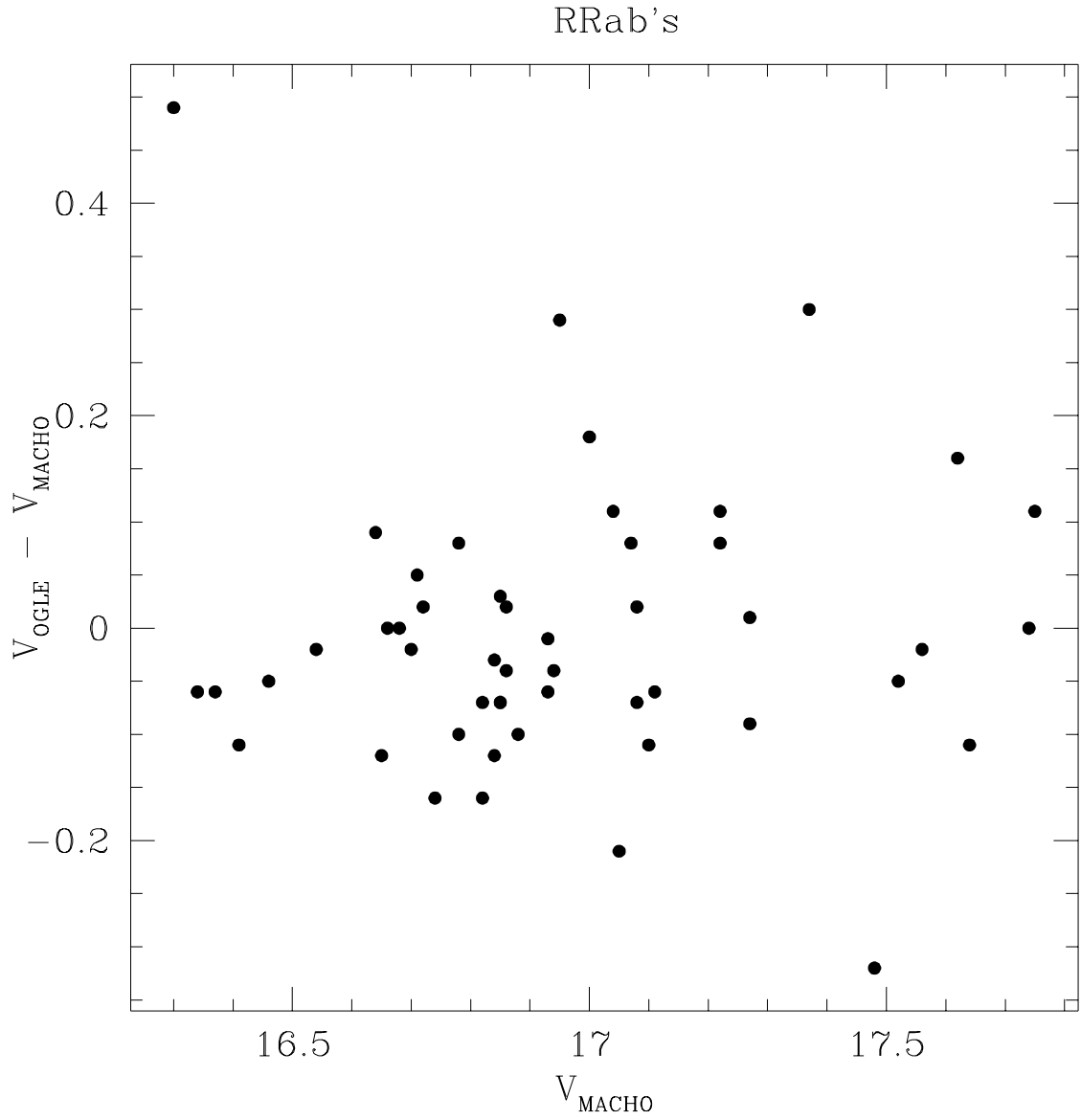
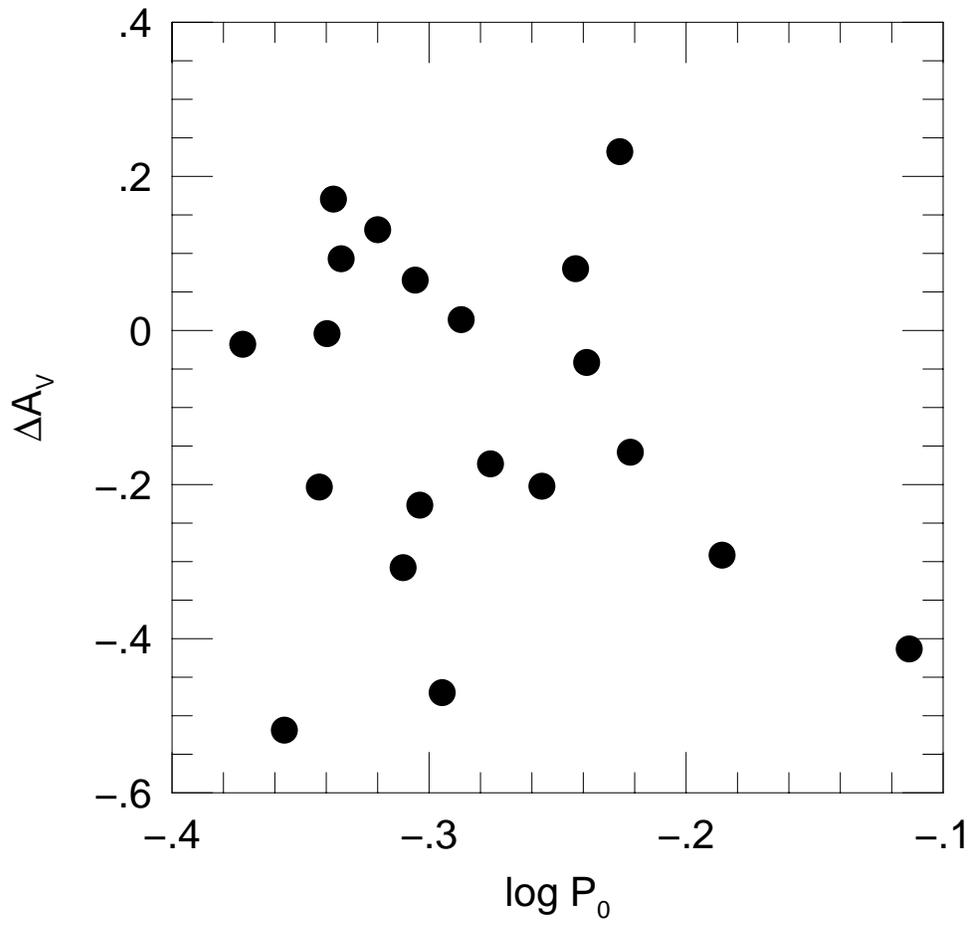
\end{document}